\RequirePackage{fix-cm}
\documentclass[twocolumn,epjc3]{svjour3}  
\smartqed  % flush right qed marks, e.g. at end of proof
\RequirePackage{graphicx}
\usepackage{atlasphysics} % Contains useful shortcuts. Uncomment to use
\usepackage{subfigure}
\usepackage{hyperref}
\journalname{Eur. Phys. J. C}
\begin{document}
\newcommand{\lumi}{34$\,$\ipb}
\newcommand{\searchlumi}{31$\,$\ipb}
\newcommand{\controllumi}{2.6$\,$\ipb}
\newcommand{\backlumi}{0.26$\,$\ipb}

\newcommand{\masslimitgeneric   }{341}
\newcommand{\masslimitregge     }{294}

\newcommand{\masslimitexpgeneric}{321}
\newcommand{\masslimitexpregge  }{263}

\hyphenation{ATLAS FORTRAN Hy-phen-a-tion}

\newcommand{\tilefrac}{$f_{\rm{Tile}}$}

%%\def\lumi{34$\,$\ipb}
%%\def\searchlumi{31$\,$\ipb}
%%\def\controllumi{2.6$\,$\ipb}
%%\def\backlumi{0.26$\,$\ipb}
%%
%%\def\masslimitgeneric   {341}
%%\def\masslimitregge     {294}
%%
%%\def\masslimitexpgeneric{321}
%%\def\masslimitexpregge  {263}
%%
%%\hyphenation{ATLAS FORTRAN Hy-phen-a-tion}
%%
%%
%%\def\tilefrac{$f_{\rm{Tile}}$}

\title{\vspace{-2.0cm}\flushleft{\normalsize\normalfont{CERN-PH-EP-2011-188}} \flushright{\vspace{-0.86cm}\normalsize\normalfont{Submitted to Eur.\ Phys.\ J.\ C}} \\[2.5cm] \flushleft{Search for Decays of Stopped, Long-Lived Particles from 7\,{\bf TeV} {\bf $pp$} Collisions with the ATLAS Detector}} 

\titlerunning{Search for Decays of Stopped, Long-Lived Particles from 7\,{\bf TeV} {\bf $pp$} Collisions with the ATLAS Detector}        % if too long for running head

\author{The ATLAS Collaboration
        %\and
        %Second Author\thanksref{e2,addr2,addr3} %etc.
}
\authorrunning{The ATLAS Collaboration} % if too long for running head

\institute{CERN, 1211 Geneva 23, Switzerland \email{atlas.publications@cern.ch}}
\date{\today}

%\date{Received: date / Accepted: date Done by editor}

\maketitle

\abstract{
New metastable massive particles with electric and colour charge are features
of many theories beyond the Standard Model.
A search is performed for long-lived gluino-based $R$-hadrons
with the ATLAS detector at the LHC using a data
sample corresponding to an integrated luminosity of $\searchlumi$. We search for evidence
of particles that have come to rest in the
ATLAS detector and decay at some later time during the periods in the LHC bunch structure
without proton-proton collisions.
No significant deviations from the expected backgrounds are observed, and a
cross-section limit is set.  It can be interpreted as excluding
gluino-based $R$-hadrons with masses less
than $\masslimitgeneric\,$GeV at the 95$\%$ C.L., for lifetimes from
10$^{-5}$ to 10$^3\,$seconds and a neutralino mass of 100$\,$GeV.
}

\section{Introduction}
\label{sec:introduction}
The search for exotic massive long-lived particles (LLPs) is an important component of
the early data exploration program of the Large Hadron Collider (LHC) experiments.
LLPs are predicted in supersymmetry (SUSY) models, such as
split-SUSY~\cite{ArkaniHamed:2004yi,Hewett:2004nw} and gauge-mediated
SUSY breaking~\cite{Mafi:1999dg}, as well as other exotic scenarios, e.g.
universal extra dimensions~\cite{Shah:2006gs}.
If LLPs carried colour charge, as predicted in a variety of Standard Model (SM)
extensions~\cite{Fairbairn:2006gg}, they would hadronise with quarks and gluons
to form colour-singlet states before interacting with the detector.
$R$-parity conserving split-SUSY presents such a model since the gluino ($\tilde{g}$)
decay---$\tilde{g}\rightarrow g \tilde{\chi}^0_1$, 
or $\tilde{g}\rightarrow q\bar{q}\tilde{\chi}^0_1$---is
suppressed by small coupling to the gravitino LSP or very large squark
($\tilde{q}$) mass\cite{Gambino:2005eh}. The possibility of
direct pair production of coloured particles through the strong nuclear force implies a large production 
cross-section at the LHC. The search described herein is particularly sensitive to such a scenario,
and results are interpreted in the context of split-SUSY.

During the hadronisation process, long-lived gluinos bind with SM quarks and gluons
from the vacuum to produce ``$R$-hadrons''\footnote{The 
term $R$-hadron has its origin in the $R$-parity quantum number in 
supersymmetry theories.}---a new heavy composite state~\cite{Farrar:1978xj}.
When $R$-hadrons scatter via the nuclear force, they can transform
their internal state by exchanging partons with the detector material 
(e.g. $\tilde{g}q\bar{q} \rightarrow \tilde{g}qqq$).
Depending on the mass hierarchy of $R$-hadron states and
the matter interaction model, these particles traverse the detector
flipping between electrically neutral, singly- and doubly-charged states.
Several properties of the $R$-hadron, such as mass and lifetime\footnote{
Lifetime refers to the gluino decays, not the $R$-hadron
transition from one hadronic state to another.}, are 
primarily determined by properties of the constituent gluino.  
Other properties, such as hadronic 
matter interactions, are dominated by the constituent quarks and gluons. 
%Since the $\tilde{g}$ would be strongly interacting, it would be directly produced
%from the initial state partons, and not often through cascade decays.
%This implies that $\tilde{g}$ and thus $R$-hadrons may be produced near
%threshold in LHC collisions, and thus move slowly in the detector, with large
%ionization energy loss.

Since $R$-hadrons may be produced near threshold in LHC collisions,
they are expected to be slow ($\beta$ significantly below 1) and have 
large ionisation energy loss.
At LHC energies, some fraction (typically several per cent) of $R$-hadrons, produced with low kinetic energy, 
would lose sufficient energy to come to rest inside the dense detector
materials of the ATLAS calorimeter.
These stopped $R$-hadrons  may have lifetimes spanning many orders of 
magnitude, and 
may decay with significant delay after the collision that created them.
It is possible to more easily detect these decays by searching for 
calorimeter energy deposits during the so-called {\it empty} 
bunch crossings~\cite{Arvanitaki:2005nq} when there is less background, 
as described in detail in Section~\ref{sec:trigger}.
Since this sample has a very low proton-proton collision rate,
the dominant backgrounds to this search are
cosmic ray muons, beam-related backgrounds and instrumental noise.
Similar searches for out-of-time decays have previously been performed by other
experiments~\cite{Abazov:2007ht,cms}.

This analysis complements previous ATLAS searches for 
long-lived particles~\cite{bib:muon_agnostic,bib:muon_LLP}
which are less sensitive to particles with initial $\beta \ll 1$.
By relying primarily on calorimetric measurements, 
this analysis is also sensitive to events where $R$-hadron charge-flipping
may make reconstruction in the inner tracker and the muon system impossible.
A potential detection of stopped $R$-hadrons could also lead to a measurement of their lifetime and decay properties. 
Moreover, the search is sensitive to any potential new physics scenario producing 
large out-of-time energy deposits in the calorimeter with minimal additional detector activity. 
The data analysed in this Letter were recorded by the ATLAS 
experiment between April and November 2010, exploiting 
proton-proton collisions at a centre-of-mass energy of 7$\,$TeV.

\section{The ATLAS detector}
\label{sec:atlas}
The ATLAS detector~\cite{ATLAS_performance} consists of an inner tracking system (ID) surrounded by a thin
superconducting solenoid, electromagnetic and hadronic calorimeters and a muon spectrometer (MS).
The ID consists of pixel and silicon microstrip detectors, surrounded by a transition radiation tracker.
The calorimeter system is based on two active media for the electromagnetic
and hadronic calorimeters: liquid argon (LAr) in the
inner and forward regions, and scintillator-tiles 
(TileCal) in the outer barrel region.  The absorber materials are lead and 
either steel, copper, or tungsten.
The calorimeters are segmented in cells which have typical size 
0.1 by 0.1 in $\eta$-$\phi$ space\footnote{ATLAS uses a right-handed coordinate system 
with its origin at the nominal interaction point (IP) in the centre of the detector and the $z$-axis along the beam 
pipe. The $x$-axis points from the IP to the centre of the LHC ring, and the $y$-axis points upward. 
Cylindrical coordinates $(r,\phi)$ are used in the transverse plane, $\phi$ being the azimuthal angle around the 
beam pipe. The pseudorapidity is defined in terms of the polar angle $\theta$ as $\eta=-\ln\tan(\theta/2)$.} in the TileCal section.
The MS, capable of reconstructing tracks within $|\eta| < 2.7$, 
uses toroidal bending fields generated by three large superconducting magnet systems.

Jets are constructed using the anti-$k_t$ jet algorithm~\cite{antikt}
with a radius parameter (in $\eta$-$\phi$ space)
set to $R = 0.4$, which assumes the energetic particles
originated from nominal interaction point.
This assumption, while incorrect, still accurately quantifies
the energy released from the stopped $R$-hadron decays occurring
in the calorimeter; the associated systematic uncertainty is discussed in
Section~\ref{sec:systematics}.
The inputs to the jet algorithm described in this paper are calorimeter energy deposits.
Jet energy is quoted without applying a hadronic calibration; only jets with energy
above 10$\,$GeV are considered in this search.
ATLAS jet reconstruction algorithms are described
in more detail elsewhere~\cite{published_jetreco}.

\section{LHC bunch structure and trigger strategy}
\label{sec:trigger}
The LHC accelerates two counter-rotating proton beams, each divided into 3564 25$\,$ns bunch slots.
When protons are injected into the LHC not every slot is filled.
In late 2010 running, slots that were filled typically had $10^{11}$ protons.
Unfilled slots could contain protons due to diffusion from filled slots,
but this was typically below $5 \times 10^8$ protons per slot.

The filled and unfilled slots can be combined to make three different
``bunch crossing'' scenarios.  A \textit{paired} crossing consists of
a filled bunch from each beam colliding in ATLAS and is by far the most likely
to produce $R$-hadrons.  An \textit{unpaired}
crossing has a filled bunch from one beam and an unfilled slot from the other.
Finally, in an \textit{empty} crossing the slots from both beams
are unfilled.  
Empty crossings typically had a proton-proton collision
rate less than one-millionth the rate in paired crossings\cite{ATLAS_lumi_determination}.
In the search sample, described in Section~\ref{sec:samples},
there are approximately 350 paired bunch crossings
organized in groups of eight.  Inside a group,
paired bunch crossings occurred every
150$\,$ns or longer, while the gaps between groups
are longer.
Most ATLAS analyses use data collected from the
paired crossings; this analysis instead searches for physics signatures
of LLPs in the empty crossings.
This is accomplished with a set of dedicated low-threshold calorimeter
triggers which may fire only in the empty or unpaired crossings where the
background to this search is much lower.

ATLAS has a three-level trigger system consisting of one hardware
and two software levels~\cite{Aad:2012xs}.
Signal candidates for this analysis are collected
using a hardware trigger requiring localised calorimeter activity 
with a 10$\,$GeV transverse energy threshold.
This trigger could fire only during an empty crossing at
least 125$\,$ns after the most recent paired bunch crossing.  
By waiting to collect data only during the empty crossings many signal
decays are lost; however, this provides a sample nearly free of all
collision backgrounds.
Although the $R$-hadrons decay at a time randomly distributed with
respect to the bunch-crossing clock (which has a 25$\,$ns period), this
does not cause any loss of efficiency as the calorimeter response is
longer than 25$\,$ns.
A sideband region to study
beam-halo muons is collected with a similar trigger that fired in the
unpaired crossings.  
Both samples are collected with only a hardware-level trigger
and without further requirement at the higher trigger levels.

\section{Simulation of $R$-hadrons}\label{sec:mc}
Monte Carlo simulations are used primarily to determine the reconstruction 
efficiency and stopping fraction of the $R$-hadrons, and to study
associated systematic uncertainties on the quantities used in the selections.
The simulated samples have gluino masses in the range 200--600$\,$GeV, 
to which the present analysis is sensitive.
Furthermore, a detailed simulation incorporating gluino generation, 
stopping and decay steps provides a modular approach for signal production.
The {\sc Pythia} program~\cite{Sjostrand:2006za} is used to simulate gluino-gluino pair production events. 
The string hadronisation model~\cite{Andersson:1983ia}, incorporating specialised hadronisation 
routines~\cite{Fairbairn:2006gg} for $R$-hadrons, is used inside {\sc Pythia}
to produce final states containing two $R$-hadrons. 

{
\begin{table}[!htbp]
\centering
\caption{Signal Monte Carlo stopping fractions for various gluino mass values with their 
statistical uncertainties.  Each efficiency is the number of $R$-hadrons that stop anywhere
in the detector  divided by the number of produced $R$-hadrons events.
Some $R$-hadrons stop in parts of the detector where there is little 
sensitivity to detect their decay.  This is accounted for in the reconstruction
efficiency in Table~\ref{tab:cutflow2jet}.
}
\begin{tabular}{l|c|c}
\hline
m$_{\tilde{g}}$ (GeV) & {\it Generic (\%)} & {\it Regge (\%)} \\
\hline
\hline
200 & 11.5$\pm$0.7 & 5.7$\pm$0.5 \\
300 & 13.1$\pm$0.7 & 4.7$\pm$0.5 \\
400 & 12.4$\pm$0.7 & 6.5$\pm$0.5 \\
500 & 13.2$\pm$0.7 & 4.5$\pm$0.5 \\
600 & 11.8$\pm$0.7 & 5.6$\pm$0.5 \\
\hline
\end{tabular}\label{tab:StoppingEfficiency}
\end{table}
}

To compensate for the fact that $R$-hadron scattering is not strongly constrained
by SM analogues, the simulation of $R$-hadron interactions in matter is
handled by a special detector response simulation~\cite{Kraan:2004tz} 
using {\sc Geant4}~\cite{Agostinelli:2002hh,sim_infra} routines based on 
two rather different scattering models with different sets of assumptions: 
the {\it Generic}~\cite{Kraan:2004tz,Mackeprang:2006gx} and {\it Regge}~\cite{deBoer:2007ii,Mackeprang:2009ad} 
models. 
Each model makes different assumptions about the $R$-hadron nuclear cross-section and
mass spectra of various internal states.
Briefly, the phenomenologies of the two models are described as follows:
\begin{description}
\item[{\it Generic}]: Limited constraints on allowed stable states permit the
occurrence of doubly charged $R$-hadrons
and a wide variety of charge-exchange scenarios.
The scattering model is purely phase space driven. 
This model is chosen as the nominal signal model for gluino $R$-hadrons.
\item[{\it Regge}]: Only one (electrically neutral) baryonic state is allowed. 
The scattering model employs a triple-{\it Regge} formalism. This is expected 
to result in a lower stopping fraction than the {\it Generic} scenario.
\end{description}
If an $R$-hadron comes to rest in ATLAS, its location is recorded.
Table~\ref{tab:StoppingEfficiency} shows the probability for an $R$-hadron
to stop as a function of the generated gluino mass 
for the two models considered.  The stopping fraction shows no significant
dependence on the gluino mass within the available simulation statistics.

These stopping locations are used as input into a second step of {\sc Pythia} where
the decays of the $R$-hadrons are simulated. 
Different models allow the gluinos to decay via 
the radiative process, 
$\tilde{g}\rightarrow g \tilde{\chi}^0_1$, 
or via $\tilde{g}\rightarrow q\bar{q}\tilde{\chi}^0_1$. 
The reconstruction efficiencies are comparable for both decay modes however,
the results are interpreted assuming a 100\% branching ratio to $g \tilde{\chi}^0_1$ in
accordance with previous results~\cite{Abazov:2007ht,cms}.
In all simulations the
neutralino mass, $m_{\tilde{\chi}^0_1}$, is fixed to 100$\,$GeV.
Interactions of the decay products with the detector are simulated with {\sc Geant4}.
These events then follow the standard ATLAS reconstruction which outputs
signal candidates.

\begin{figure}[htbp]
 \begin{center}
   \includegraphics[width=0.48\textwidth]{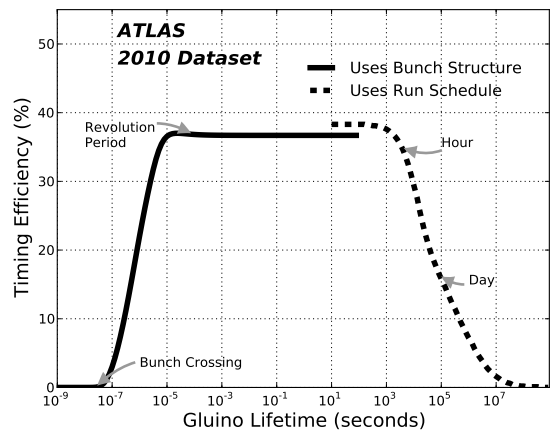}
 \caption{
Percentage timing efficiency as a function of gluino lifetime.
The region $10^{-5}$ to $10^3$$\,$seconds has the highest
efficiency, approximately 37$\%$.  For shorter lifetimes many $R$-hadrons
decay in paired bunch crossings while for longer lifetimes many decay when ATLAS
is not taking data.  The solid line is calculated with the assumption
that no $R$-hadron survived from one run to the next (allowing a per run analysis for
shorter lifetimes).  The dashed line is calculated after averaging over bunch structure
but does allow $R$-hadrons to decay in a run they are not produced in.
 \label{fig:timing}}
 \end{center}
\end{figure}

An additional inefficiency arises because only a fraction of the
stopped $R$-hadrons decay in an empty crossing, while ATLAS is taking data.
This inefficiency is a function of the gluino lifetime, and is
calculated from a detailed knowledge of the bunch structure across different
LHC fills (relevant for short lifetimes) and the ATLAS data acquisition
schedule (relevant for long lifetimes).  The timing efficiency versus
lifetime is shown in Figure~\ref{fig:timing}; the plateau corresponds
to 37\% efficiency for lifetimes from $10^{-5}$ to $10^3\,$ seconds.
These lifetimes are long enough to survive until an isolated empty crossing
which may be many microseconds after the production event.  However,
these lifetimes are short enough so that most of the $R$-hadrons do not decay
outside of an ATLAS data-taking run, which is typically ten to twenty hours.

\section{Data samples}\label{sec:samples}
As discussed in Section~\ref{sec:trigger}, data are collected with
calorimeter-based triggers in the empty and unpaired crossings.
To construct two sidebands and a signal search sample, the data from empty bunch
crossings are separated into early, middle and late 2010 subsets, respectively.
The total integrated luminosity in paired crossings concurrent with these three empty-crossing
samples is $\lumi$ \cite{ATLAS_lumi_determination}.
Each sample has a different signal to background ratio due to the
rapidly increasing instantaneous luminosity of the LHC during 2010.

To predict the number of cosmic ray muons---the dominant background---both the
amount of time and varying number of empty bunches per beam revolution
in each data-taking run must be accounted for.
The trigger live-time (in units of crossing-hours) is quantified
by taking the product of the number of hours the trigger is active and the
number of empty bunch crossings per beam revolution in that LHC filling scheme.

The first data sample, the {\it background sample}, had low beam current and
corresponds to a
trigger live-time of 1.07$\times$10$^{6}$ crossing-hours and $\backlumi$
of integrated luminosity; it is used to estimate the background from cosmic ray muons.
A second sample, the {\it control sample}, corresponds to 0.55$\times$10$^{6}$
crossing-hours, and an integrated luminosity of $\controllumi$.  This sample
has different beam conditions, and acts as a cross-check
for the cosmic ray muon background estimate.
The signal search is performed in the {\it search sample}, which corresponds to
0.32$\times$10$^{6}$ crossing-hours.
Data collected in the search sample reflect an integrated luminosity of
$\searchlumi$ with a peak instantaneous luminosity of approximately
$2.1\times 10^{32}\,$cm$^{-2}\,$s$^{-1}$.
The probability of producing an $R$-hadron in either the background or control
sample is small compared to the search sample.

Finally a small sample from the unpaired crossings is used
to quantify the background induced by beam-halo muons.

\section{Candidate selection}\label{sec:selection}
In addition to the signal candidates, several background processes survive
the trigger: beam-halo muons,
proton-proton collisions, proton-gas collisions, calorimeter noise and cosmic ray muons.
The first three of these processes arise from the non-zero proton population
in the empty bunches, as described in Section~\ref{sec:trigger}.
Beam-halo muons are produced when a stray $3.5\,$TeV proton 
collides with a collimator or residual gas many metres
away from ATLAS.  The subsequent shower decays or is blocked by shielding, 
but occasionally a muon survives to traverse the detector.

The search and sideband samples are divided into two exclusive channels
based on the number of reconstructed jets with energy over 10$\,$GeV,
$N_{\rm{jets}}$.
Lower energy $R$-hadron decays tend to be reconstructed as a single jet,
but very energetic ones may be reconstructed as multiple jets due to their
larger spatial extent.
The two channels, denoted single jet and multi-jet, differ primarily in their
respective noise rejection criteria, labelled as ``jet cleaning''.
Without requirements on a second jet to reject background, the single jet
channel, $N_{\rm{jets}}=1$, uses a more demanding calorimeter selection.
Selection criteria described below that are not specifically mentioned in the
event yield summary (Table~\ref{tab:cutflow2jet})
fall under the jet cleaning label for either the single or multi-jet channel.

\begin{table*}
\begin{center}
    \caption{
Event yield at various levels of the selection for the
single jet (top) and multi-jet (bottom) channels.
The yields of the background and control samples are
scaled by detector live-time to the search sample. 
The background and control samples agree well in the bottom three rows for both 
channels, showing the live-time scaling works after detector noise is removed 
(via jet cleaning).
Agreement is not expected before jet cleaning
due to the sporadic nature of calorimeter noise across 2010.
Since the rate of beam-halo muons scales with beam current it
only contributes significantly to the search sample as
discussed in Section~\ref{sec:background}.
Only statistical uncertainties are shown in the table.
%and are also scaled by the live-time (which
%is assumed here to have no uncertainty).
    \label{tab:cutflow2jet}}
\begin{tabular}{l|c|c|c}
 \hline
Selection criterion & Background sample & Control sample & Search sample \\
\hline\hline

Data Quality                &1119200 $\pm$ 600  &1038600 $\pm$ 800  &1579949 \\
$|$Jet $\eta|\:<\:$1.2      &875600  $\pm$ 500  &739600  $\pm$ 700  &1094196 \\
$N_{\rm{jets}} = $ 1        &71800   $\pm$ 200  &72800   $\pm$ 200  &1089374 \\
Jet Cleaning                &5860    $\pm$ 40   &5860    $\pm$ 60   &5615    \\
Muon Segment Veto           &4       $\pm$ 1    &5       $\pm$ 2    &9       \\
Jet Energy $>$ 100 GeV      &0.3     $\pm$ 0.3  &0.6     $\pm$ 0.6  &0       \\
\hline
Data Quality                &1119200 $\pm$ 600  &1038600 $\pm$ 800  &1579949    \\ 
$|$Jet $\eta|\:<\:$2.2      &1093300 $\pm$ 600  &1012400 $\pm$ 800  &1507017    \\ 
1$\:<\: N_{\rm{jets}}\:<$10 &22480   $\pm$ 80   &22700   $\pm$ 100  &24902      \\ 
Jet Cleaning                &8650    $\pm$ 50   &8310    $\pm$ 70   &8036       \\ 
Muon Segment Veto           &1       $\pm$ 0.6  &4       $\pm$ 2    &3          \\  
Jet Energy $>$ 100 GeV      &0.6     $\pm$ 0.4  &0.6     $\pm$ 0.6  &1          \\  
\hline

\end{tabular}
\end{center}
\end{table*}

\subsection{Criteria common to both channels}
All events in the three empty-crossing data samples, described 
in Section~\ref{sec:samples}, must pass basic beam, detector and
data quality requirements.
%All events in the three empty-crossing data samples, described in Section~\ref{sec:samples},
%must pass basic beam, detector and data quality requirements; the
%inefficiency from these requirements is accounted for during the results interpretation.
Events with proton-proton collisions or proton-gas collisions, 
which happen at a very low rate in empty crossings, are rejected
by requiring no reconstructed tracks in the ID.
Additionally, events where the missing transverse momentum is less than half of the
leading jet transverse momentum are vetoed.
The criterion on missing transverse momentum removes the rare
collision events since they tend to produce jets with balanced transverse momentum.
A signal event, however, produces only one
localised energy deposit, since even if both $R$-hadrons stop in
the detector they are likely to decay at times separated by more than 25$\,$ns.
The trigger system would identify these decays as separate events
because its timing resolution is significantly smaller than 25$\,$ns.
%Since the trigger system's timing resolution is significantly smaller than
%25$\,$ns, it identifies them as separate events.

To reduce the contribution from calorimeter noise, the most energetic jet
must have an energy greater than 50$\,$GeV, and be in the central region ($|\eta| < 2.2$).
%The selection at 50$\,$GeV is defined to veto calorimeter noise so that
%data sample agreement can be confirmed; it is superseded by a 100$\,$GeV selection
%for the final search region.
Furthermore, $n_{90}$, the minimum number of calorimeter cells that collectively
contain 90$\%$ of the jet energy, must be greater than three.

Cosmic ray and beam-halo muons are vetoed by requiring the absence
of a muon segment anywhere in the MS.  A muon segment is formed
when there is correlated activity in a single detector station of the MS,
and serves as input to more sophisticated track finding algorithms.
This presents the common selection in the different jet channels and 
is used to study the data sample. In the final search only events with
leading jet energy over 100$\,$GeV are considered.
%Finally, only events with leading jet energy over 100$\,$GeV are considered.
This threshold optimises the search reach while maintaining sensitivity to
lighter gluino models.

\subsection{Single jet criteria}
To remove effects from the noisier calorimeter endcaps, the leading
jet must be well within the barrel region ($|\eta| < 1.2$).
Jets must have at least half of their energy
in the TileCal (\tilefrac $\;> 50\%$) to reject beam-halo muons, which occur
more frequently closer to the beam pipe.
The energy weighted transverse size of the jet, in
$\eta\!-\!\phi$ space, must be greater than 0.04.

\subsection{Multi-jet criteria}
The requirement of a second jet in the multi-jet channel,
$1 < N_{\rm{jets}} < 10$, strongly suppresses the noise and beam-halo
contributions, allowing for looser jet cleaning cuts.  
The energy of the sub-leading jet must be
at least 15$\,$GeV to ensure it is well reconstructed;
requiring $N_{\rm{jets}} < 10$
vetoes several events with bursts of calorimeter noise.
In this channel,
there is no jet transverse size requirement; however, one of the two leading jets
must have \tilefrac $\;> 10\%$.

\subsection{Data sample consistency}
\begin{figure}[htbp]
  \subfigure[Jet energy for single jet channel]{\label{fig:jete_onejet}
   \includegraphics[width=0.48\textwidth]{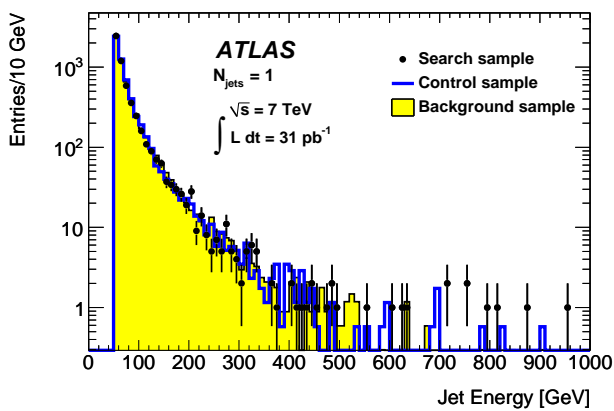}}
  \subfigure[Leading jet energy for multi-jet channel]{\label{fig:jete_twojet}
    \includegraphics[width=0.48\textwidth]{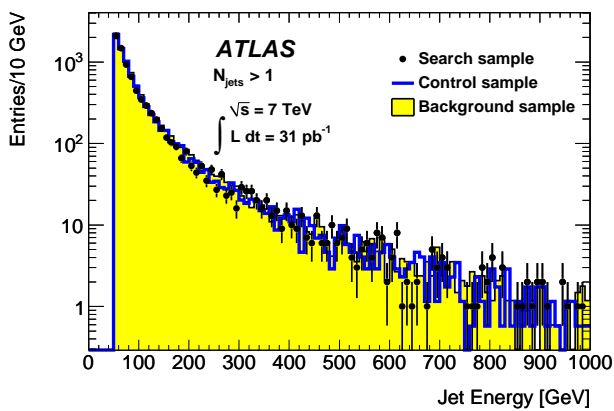}}
\caption{
Distributions of jet energy for the single jet (a) and multi-jet (b) channel. 
Events are plotted for the three empty-crossing data samples
considered in the analysis, scaled to the detector live-time of the search sample, 
prior to applying the muon segment veto.  Cosmic ray muons, whose rate
is independent of LHC luminosity, dominate
the data samples before the muon segment veto is applied.} 
\label{fig:jetplots}
\end{figure}

A very small fraction of the cosmic ray muons incident on the ATLAS detector 
deposit sufficient energy in the calorimeters to mimic the $R$-hadron signal.
This is the dominant background in all data samples prior to applying the muon segment veto.
Figures~\ref{fig:jete_onejet} and~\ref{fig:jete_twojet} show the jet energy
distributions, scaled to live-time, of the events selected in the single
and multi-jet channels, respectively.  The distributions show the event
populations after applying the jet cleaning criteria but prior
to the muon segment veto.
Agreement among the three empty-crossing data samples is found, demonstrating
that cosmic ray muons dominate the background at this point.

\subsection{Signal efficiency}
\label{sec:sigeff}
The same selection criteria are applied to both the data 
and the signal sample; their affect on signal
efficiency is calculated using the simulation.
%The efficiency for an $R$-hadron decay event to pass these criteria is calculated
%using the simulation.
There is an
additional small signal loss when a muon segment is reconstructed
in the same event as the signal decay.  
Muon segments due to cosmic rays or noise would cause the event to be vetoed
but they are not included in the simulation, unlike those due to signal
decays. Using data from the empty bunches acquired with a random trigger,
it is estimated that 93\% of decays would pass the muon segment veto and
remain in the search sample.

The reconstruction efficiency is modified by the different stopping models,
since they affect the stopping locations of the $R$-hadrons in the calorimeter.
Both channels' signal reconstruction efficiencies are summarised in Table~\ref{tab:signalEfficiency}.
Very energetic $R$-hadron decays are more likely to be reconstructed
as several jets, due to their large spatial extent.  This causes
the single jet channel efficiency to drop slightly, while the multi-jet
channel efficiency grows, for heavier $R$-hadrons.

\begin{table*}[!htbp]
\centering
\caption{Signal reconstruction efficiency for various gluino mass 
values with a fixed $\tilde{\chi}^0_1$ mass of 100$\,$GeV. 
Efficiencies are shown for the single jet (multi-jet) channels and the uncertainties quoted are statistical only.
Only the process $\tilde{g}\rightarrow g \tilde{\chi}^0_1$ is considered here.
These values do not take into account stopping inefficiencies,
timing inefficiency, or losses from random muon segments discussed in 
Section~\ref{sec:mc} and Section~\ref{sec:sigeff}.
}
\begin{tabular}{l|c|c}
\hline
m$_{\tilde{g}}$ (GeV) & {\it Generic (\%)} & {\it Regge (\%)} \\ 
\hline
\hline
200 & 0.9$\pm$0.3 (0.1$\pm$0.1)  & 1.6$\pm$0.3  (0.0$\pm$0.0)  \\
300 & 8.8$\pm$0.6 (4.1$\pm$0.4)  & 10.3$\pm$0.8 (6.1$\pm$0.6)  \\
400 & 9.5$\pm$0.8 (6.5$\pm$0.7)  & 13.1$\pm$0.9 (8.9$\pm$0.7)  \\
500 & 8.6$\pm$0.8 (9.0$\pm$0.8)  & 9.9$\pm$0.8  (12.0$\pm$0.8) \\
600 & 8.3$\pm$0.7 (9.8$\pm$0.8)  & 9.8$\pm$0.7  (14.1$\pm$0.8) \\
\hline
\end{tabular}\label{tab:signalEfficiency}
\end{table*}

\begin{table*}[!htbp]
\begin{center}
    \caption{The number of observed and expected events for each channel after the muon
    segment veto and jet energy requirement are applied, as in Table~\ref{tab:cutflow2jet}. 
    The cosmic ray muon and beam-halo expectations are derived from the background sample and
    unpaired crossings, respectively.
    Only statistical uncertainties are shown in the table.
    %The uncertainties in the table are statistical only.
    %and are also scaled by the appropriate factor.
    }
    \begin{tabular}{c|ccc}
    \hline\hline
Selection   &\multicolumn{3}{|c}{Number of events}     \\
criterion   &  Expected cosmic ray muons& Expected beam-halo  & Observed  \\
    \hline

Single Jet: Muon Veto   & $4    \pm 1   $ & $4    \pm 1.8 $ & 9   \\
Single Jet: $>$ 100$\,$GeV & $0.3  \pm 0.3 $ & $0.5  \pm 0.2 $ & 0   \\
Multi-Jet:  Muon Veto   & $1    \pm 0.6 $ & $0.8  \pm 0.8 $ & 3   \\
Multi-Jet:  $>$ 100$\,$GeV & $0.6  \pm 0.4 $ & $0.2  \pm 0.2 $ & 1   \\

\hline
    \end{tabular}\label{tab:yields}
\end{center}
\end{table*}

\section{Systematic uncertainties on the signal}
\label{sec:systematics}
A variety of systematic uncertainties are investigated; their descriptions and magnitudes
are given in the following list.
Adding these uncertainties in quadrature yields a 
total systematic uncertainty of 23\% on the predicted number of signal events. 

\begin{itemize}
    \item $R$-hadron Nuclear Interaction ($17\%$):
        To account for theoretical uncertainties in the interaction model,
        the cross-section for the quarks in the $R$-hadron
        to interact with a nucleus is varied up and down by a factor of two.
        The stopping fraction is recalculated and the largest
        deviation from the nominal value is taken as the
        uncertainty.  The anticorrelation between the reconstruction efficiency
        and the stopping efficiency is conservatively not taken into account.
        This procedure is applied for both stopping models, and the larger deviation is quoted.
    \item Selection Criteria ($9.9\%$):
        Each individual selection criterion is varied by $\pm10\%$ except
        for the $n_{90}$ requirement which is varied by $\pm 1\,$cell.
        The largest difference across all simulated signal samples and selection flow is
        used as the systematic uncertainty.
        The $10\%$ variation serves as
        a conservative estimate 
        of the effects of jet energy scale and resolution uncertainties and other detector effects.
        This variation includes effects arising from the jets originating inside
        the calorimeter and propagating in unusual directions.
        This number also accounts for the uncertainty from the limited 
        Monte Carlo statistics.
    \item Luminosity ($3.4\%$):
        This uncertainty is taken directly from Ref.~\cite{ATLAS_lumi}.
    \item Calorimeter Timing ($3\%$):
        In the signal simulation, each $R$-hadron decay is given a random timing 
        offset between $\pm 50\,$ns relative to the nominal bunch crossing time.
        This allows the quantification of the calorimeter response to 
        significantly out-of-time decays.
        %Since the $R$-hadron decay triggers the detector readout, 
        In reality, a signal
        decay will cause the trigger to be fired in the first 25$\,$ns window in
        which it deposits enough transverse energy to satisfy the calorimeter
        trigger threshold (10$\,$GeV);
%        In reality, if a decay happened more than one or two
%        nanoseconds out of the 25$\,$ns window that defines a bunch crossing, it would
%        likely be triggered in and attributed to a different bunch crossing;
        the simulation, however, only models the trigger in one
        bunch crossing.
        To take this into account, the signal
        efficiency is measured in the following windows
        [--10, 15], [--5, 20] and [0, 25]$\,$ns relative to the nominal
        bunch crossing time.  A $3\%$ fractional variation in
        the efficiency is observed.  The window is conservatively varied by 5$\,$ns 
        since the 
        calorimeter channel-to-channel timing uniformity
        is of order 2$\,$ns.

    \item Production Cross-Section ($10\%$): The same procedure 
        as in Ref.~\cite{bib:muon_agnostic} is used. The production
        cross-section from {\sc Prospino~\cite{Beenakker:1996ch}} is calculated using the sparticle mass
        as the renormalisation scale with uncertainties estimated by varying
        the renormalisation and factorisation scales upward and downward
        by a factor of two.
\end{itemize}

\section{Background estimation}
\label{sec:background}
To quantify the expected number of background events in the search sample we
investigate three different sources: cosmic ray muons, beam-halo muons,
and calorimeter noise.  Since cosmic ray muons occur at a constant rate regardless
of LHC conditions their rate is measured using the background sample.  
From this rate, $0.3\pm0.3$ ($0.6\pm0.4$) events are expected to pass the single
(multi-) jet selection in the search sample, as shown in Table~\ref{tab:cutflow2jet}.
Here the uncertainties are statistical only.

The rate of beam-halo muons changes with the LHC conditions, most importantly
with the beam current.
Beam-halo muons should contribute significantly only to the search sample
where the beam current is far higher than for the other data samples.
The number of beam-halo muons expected to pass the selection is calculated using
data collected in the unpaired crossings from the same LHC fills as
the empty crossing data.  These events have a filled bunch
in only one beam and, after requiring a far forward muon segment ($|\eta| > 2.2$),
correspond to a nearly pure beam-halo sample.  Of
the beam-halo events that passed the single (multi-) jet criteria, 
all but $0.6\%$ ($0.05\%$) leave a muon segment.  
In the empty bunches there are 83 (531) events with far forward
muon segments consistent with beam-halo
that otherwise pass the selection.
Thus 0.5$\pm$0.2 (0.2$\pm$0.2) beam-halo events are expected to pass
the selection criteria without leaving a muon segment in the search sample.
This expected yield of beam-halo background is added to the final numbers in the background
sample of Table~\ref{tab:cutflow2jet} to obtain the overall 
expected background yield shown in Table~\ref{tab:yields}. 
After the jet cleaning selection much less than 1 event is expected from
calorimeter noise in either of the search channels.

\section{Results}
\label{sec:results}
\begin{figure}[htbp]
  \subfigure[Jet energy for single jet channel]{\label{fig:jete_onejet_sig}
   \includegraphics[width=0.48\textwidth]{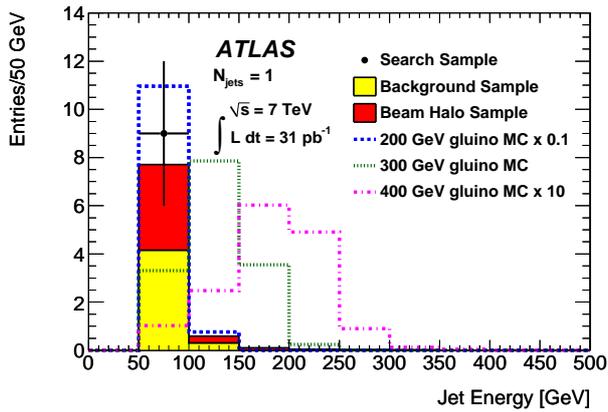}}
  \subfigure[Leading jet energy for multi-jet channel]{\label{fig:jete_twojet_sig}
    \includegraphics[width=0.48\textwidth]{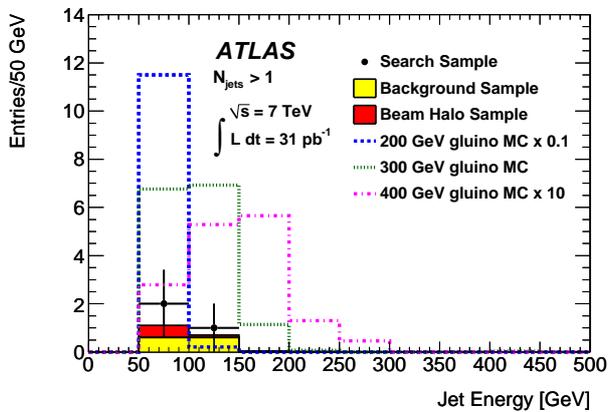}}
\caption{Distributions of jet energy for the single jet (a) and multi-jet (b)
channels, for selected events before the leading jet energy requirement
is applied. In each case the signal region is defined as those events
which have leading jet energy above 100 GeV. The background and
control samples predict cosmic-ray muon yields, but do not include
the background from beam halo. The signal distributions shown correspond
to gluino decays to a gluon and a \nino with 100$\,$GeV mass.
The signal distributions do not change for any gluino $R$-hadron lifetime
between $10^{-5}$ and $10^3\,$seconds.
}
\label{fig:jetenergy_sigRegions}
\end{figure}

The leading jet energy distribution of the selected events is shown
in Figures~\ref{fig:jete_onejet_sig} and~\ref{fig:jete_twojet_sig}
for the single jet and multi-jet channels respectively. No excess of events is
observed in either signal region beyond leading
jet energies of 100$\,$GeV.

Limits are set for the single and multi-jet 
measurements separately; however, since the single jet channel 
gives an expected limit on the cross-section which is half that of the multi-jet
channel, the observed limit from the single jet channel is quoted as the final result.
This limit applies to $R$-hadrons decaying to a gluon and 
lightest supersymmetric particle ($m_{\tilde{\chi}^0_1}$=100$\,$GeV)
with a lifetime between $10^{-5}$ to $10^3$ seconds as described in 
Section~\ref{sec:mc}.
The limit takes into account systematic uncertainties on the signal 
cross-section and efficiency as well as on the 
background estimate, as described above.

\begin{figure}[htbp]
 \includegraphics[width=0.48\textwidth]{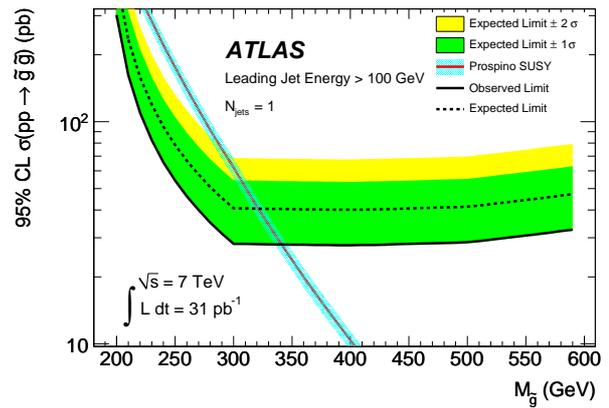}
 \caption{The 95\% C.L. limit on the gluino pair production cross-section 
        as a function of gluino mass, $m_{\tilde{g}}$, 
        assuming $m_{\tilde{\chi}^0_1}$=100$\,$GeV.
        Limits are shown for the {\it Generic} stopping model and $\tilde{g} \rightarrow g\nino$ decay.}
\label{fig:mass_limits}
\end{figure}

Limits are set on the signal cross-section for each gluino mass,
$m_{\mathrm{\tilde{g}}}$, using a Bayesian method~\cite{bayesian} with a uniform prior.
Given the expected cross-section as a function of mass, 
and a limit on the expected number of signal events, 
gluino $R$-hadrons with a mass $m_{\mathrm{\tilde{g}}} < \masslimitgeneric\,\mathrm{GeV}$
%gluino $R$-hadrons with a mass \mbox{$m_{\mathrm{\tilde{g}}} < \masslimitgeneric\,$GeV} 
($\masslimitregge\,$GeV) are excluded
at 95$\%$ credibility level (C.L.) for the {\it Generic} ({\it Regge}) model.
Gluino masses $m_{\mathrm{\tilde{g}}} < 200\,$GeV are not investigated.
Figure~\ref{fig:mass_limits}
shows the expected and observed 95$\%$ C.L. limit and the $\pm$1 and $\pm$2
standard deviation bands.
The cross-section limit changes rapidly in the
\mbox{$200 < m_{\mathrm{\tilde{g}}} < 300\,$GeV} region as, with the 
increasing gluino mass, more decays pass the $100\,$GeV jet energy selection.
Above $300\,$GeV the limit depends only moderately on the
gluino mass.
The $-2\sigma$ and $-1\sigma$ variations of the expected limit coincide
with the observed limit since no events are observed in the final
selected data sample, and the background is low.

Both the D0 and CMS collaborations performed searches yielding comparable results
for similar lifetime ranges.
The D0 collaboration analysed 410$\,$\ipb of proton-antiproton data collected
from collisions at $\sqrt{s} = 1.96\,\mathrm{TeV}$ and derived a 95\% C.L. limit,
excluding $m_{\tilde{g}} < 270\,$GeV.
The CMS collaboration used 10$\,$\ipb of proton-proton data collected from
collisions at $\sqrt{s} = 7\,\mathrm{TeV}$.  They derived a 95\% C.L. limit,
excluding $m_{\tilde{g}} < 370\,$GeV.
%The CMS analysis used both a larger number of slot crossings to search in,
%as well as a larger calculated stopping fraction than the analysis presented
%here.

%Both the D0 and CMS collaborations performed searches yielding comparable results
%for similar lifetime ranges.
%The D0 collaboration analysed 410$\,$\ipb of proton-antiproton data collected
%from collisions at $\sqrt{s} = 1.96\,\mathrm{TeV}$ and derived a 95\% C.L. limit
%on production cross-section of approximately 1$\,$pb, excluding $m_{\tilde{g}} < 270\,$GeV.
%The CMS collaboration used 10$\,$\ipb of proton-proton data collected from
%collisions at $\sqrt{s} = 7\,\mathrm{TeV}$.  They derived a 95\% C.L. limit
%on the cross-section of approximately 10$\,$pb, excluding $m_{\tilde{g}} < 370\,$GeV.
%The CMS analysis used both a larger number of slot crossings to search in,
%as well as a larger calculated stopping fraction than the analysis presented
%here.

\section{Summary}
\label{sec:summary}
A search is presented for long-lived gluinos which have stopped in the 
ATLAS detector, using 7$\,$TeV proton-proton 
collisions at the LHC. No evidence for the subsequent 
decay of these particles into $g\nino$ is found, 
in a dataset with peak instantaneous luminosity of $2\times 10^{32}\,\mathrm{cm}^{-2}\,\mathrm{s}^{-1}$
%in a dataset with peak instantaneous luminosity of \mbox{$2\times 10^{32}\,$cm$^{-2}\,$s$^{-1}$}
and an integrated luminosity of $\searchlumi$, during
time periods where there are no proton-proton collisions. 
A dedicated calorimeter trigger is employed, and
the observed events are consistent with the background expectation derived
from control samples in data.
Limits on the
gluino pair production as a function of gluino lifetime are derived and exclude 
$200 < m_{\tilde{g}} < \masslimitgeneric\,$GeV at 95$\%$ C.L. 
for lifetimes between $10^{-5}$ to $10^3$ seconds with a fixed neutralino mass 
of $m_{\tilde{\chi}^0_1}$=100$\,$GeV and the {\it Generic} matter interaction model.

\begin{acknowledgements}
% Acknowledgements for papers with collision data
% Version 6-Jan-2012

We thank CERN for the very successful operation of the LHC, as well as the
support staff from our institutions without whom ATLAS could not be
operated efficiently.

We acknowledge the support of ANPCyT, Argentina; YerPhI, Armenia; ARC,
Australia; BMWF, Austria; ANAS, Azerbaijan; SSTC, Belarus; CNPq and FAPESP,
Brazil; NSERC, NRC and CFI, Canada; CERN; CONICYT, Chile; CAS, MOST and NSFC,
China; COLCIENCIAS, Colombia; MSMT CR, MPO CR and VSC CR, Czech Republic;
DNRF, DNSRC and Lundbeck Foundation, Denmark; ARTEMIS and ERC, European Union;
IN2P3-CNRS, CEA-DSM/IRFU, France; GNAS, Georgia; BMBF, DFG, HGF, MPG and AvH
Foundation, Germany; GSRT, Greece; ISF, MINERVA, GIF, DIP and Benoziyo Center,
Israel; INFN, Italy; MEXT and JSPS, Japan; CNRST, Morocco; FOM and NWO,
Netherlands; RCN, Norway; MNiSW, Poland; GRICES and FCT, Portugal; MERYS
(MECTS), Romania; MES of Russia and ROSATOM, Russian Federation; JINR; MSTD,
Serbia; MSSR, Slovakia; ARRS and MVZT, Slovenia; DST/NRF, South Africa;
MICINN, Spain; SRC and Wallenberg Foundation, Sweden; SER, SNSF and Cantons of
Bern and Geneva, Switzerland; NSC, Taiwan; TAEK, Turkey; STFC, the Royal
Society and Leverhulme Trust, United Kingdom; DOE and NSF, United States of
America.

The crucial computing support from all WLCG partners is acknowledged
gratefully, in particular from CERN and the ATLAS Tier-1 facilities at
TRIUMF (Canada), NDGF (Denmark, Norway, Sweden), CC-IN2P3 (France),
KIT/GridKA (Germany), INFN-CNAF (Italy), NL-T1 (Netherlands), PIC (Spain),
ASGC (Taiwan), RAL (UK) and BNL (USA) and in the Tier-2 facilities
worldwide.

\end{acknowledgements}

\bibliographystyle{spphys}       % APS-like style for physics
\bibliography{maintext}

\begin{thebibliography}{10}
\providecommand{\url}[1]{{#1}}
\providecommand{\urlprefix}{URL }
\expandafter\ifx\csname urlstyle\endcsname\relax
  \providecommand{\doi}[1]{DOI \discretionary{}{}{}#1}\else
  \providecommand{\doi}{DOI \discretionary{}{}{}\begingroup
  \urlstyle{rm}\Url}\fi

\bibitem{ArkaniHamed:2004yi}
N.~Arkani-Hamed, S.~Dimopoulos, G.F. Giudice, A.~Romanino, Nucl. Phys.
  \textbf{B709}, 3 (2005).
\newblock \doi{10.1016/j.nuclphysb.2004.12.026}

\bibitem{Hewett:2004nw}
J.L. Hewett, B.~Lillie, M.~Masip, T.G. Rizzo, JHEP \textbf{09}, 070 (2004).
\newblock \doi{10.1088/1126-6708/2004/09/070}

\bibitem{Mafi:1999dg}
A.~Mafi, S.~Raby, Phys. Rev. \textbf{D62}, 035003 (2000).
\newblock \doi{10.1103/PhysRevD.62.035003}

\bibitem{Shah:2006gs}
N.R. Shah, C.E.M. Wagner, Phys. Rev. \textbf{D74}, 104008 (2006).
\newblock \doi{10.1103/PhysRevD.74.104008}

\bibitem{Fairbairn:2006gg}
M.~Fairbairn, et~al., Phys. Rept. \textbf{438}, 1 (2007).
\newblock \doi{10.1016/j.physrep.2006.10.002}

\bibitem{Gambino:2005eh}
P.~Gambino, G.~Giudice, P.~Slavich, Nucl.Phys. \textbf{B726}, 35 (2005).
\newblock \doi{10.1016/j.nuclphysb.2005.08.011}

\bibitem{Farrar:1978xj}
G.R. Farrar, P.~Fayet, Phys. Lett. \textbf{B76}, 575 (1978).
\newblock \doi{10.1016/0370-2693(78)90858-4}

\bibitem{Arvanitaki:2005nq}
A.~Arvanitaki, S.~Dimopoulos, A.~Pierce, S.~Rajendran, J.G. Wacker, Phys. Rev.
  \textbf{D76}, 055007 (2007).
\newblock \doi{10.1103/PhysRevD.76.055007}

\bibitem{Abazov:2007ht}
V.M. Abazov, et~al., Phys. Rev. Lett. \textbf{99}, 131801 (2007).
\newblock \doi{10.1103/PhysRevLett.99.131801}

\bibitem{cms}
{CMS Collaboration}, Phys. Rev. Lett. \textbf{106}, 011801 (2011)

\bibitem{bib:muon_agnostic}
{ATLAS Collaboration}, Phys. Lett. \textbf{B701}, 1 (2011)

\bibitem{bib:muon_LLP}
{ATLAS Collaboration}, Phys. Lett. \textbf{B703}, 428 (2011)

\bibitem{ATLAS_performance}
{ATLAS Collaboration}, JINST \textbf{3}, S08003 (2008)

\bibitem{antikt}
{M.~Cacciari, G.~P.~Salam, and G.~Soyez}, JHEP \textbf{04}, 063 (2008)

\bibitem{published_jetreco}
{ATLAS Collaboration}, Eur. Phys. J. \textbf{C71}, 1521 (2011).
\newblock \doi{10.1140/epjc/s10052-010-1512-2}

\bibitem{ATLAS_lumi_determination}
{ATLAS Collaboration}, Eur. Phys. J. \textbf{C71}, 1630 (2011).
\newblock \doi{10.1140/epjc/s10052-011-1630-5}

\bibitem{Aad:2012xs}
A.~Collaboration, Eur.Phys.J. \textbf{C72}, 1849 (2012).
\newblock \doi{10.1140/epjc/s10052-011-1849-1}

\bibitem{Sjostrand:2006za}
T.~Sjostrand, S.~Mrenna, P.~Skands, JHEP \textbf{05}, 026 (2006)

\bibitem{Andersson:1983ia}
B.~Andersson, G.~Gustafson, G.~Ingelman, T.~Sjostrand, Phys. Rept. \textbf{97},
  31 (1983).
\newblock \doi{10.1016/0370-1573(83)90080-7}

\bibitem{Kraan:2004tz}
A.C. Kraan, Eur. Phys. J. \textbf{C37}, 91 (2004).
\newblock \doi{10.1140/epjc/s2004-01997-7}

\bibitem{Agostinelli:2002hh}
S.~Agostinelli, et~al., Nucl. Instrum. Meth. \textbf{A506}, 250 (2003).
\newblock \doi{10.1016/S0168-9002(03)01368-8}

\bibitem{sim_infra}
{ATLAS Collaboration}, Eur. Phys. J. \textbf{C70}, 823 (2010).
\newblock \doi{10.1140/epjc/s10052-010-1429-9}

\bibitem{Mackeprang:2006gx}
R.~Mackeprang, A.~Rizzi, Eur. Phys. J. \textbf{C50}, 353 (2007).
\newblock \doi{10.1140/epjc/s10052-007-0252-4}

\bibitem{deBoer:2007ii}
Y.R. de~Boer, A.B. Kaidalov, D.A. Milstead, O.I. Piskounova, J. Phys.
  \textbf{G35}, 075009 (2008).
\newblock \doi{10.1088/0954-3899/35/7/075009}

\bibitem{Mackeprang:2009ad}
R.~Mackeprang, D.~Milstead, Eur. Phys. J. \textbf{C66}, 493 (2010).
\newblock \doi{10.1140/epjc/s10052-010-1262-1}

\bibitem{ATLAS_lumi}
{ATLAS Collaboration}, ATLAS-CONF-2011-011

\bibitem{Beenakker:1996ch}
W.~Beenakker, R.~Hopker, M.~Spira, P.M. Zerwas, Nucl. Phys. \textbf{B492}, 51
  (1997).
\newblock \doi{10.1016/S0550-3213(97)00084-9}

\bibitem{bayesian}
J.~Heinrich, L.~Lyons, {Annual review of Nuclear and Particle Science}
  \textbf{57}, 145 (2007)

\end{thebibliography}

\clearpage
\onecolumn
% ATLAS Collaboration author list for 01-AUG-2011
% Data extracted on 16-Mar-2012 for paperid 123
%\documentclass[11pt]{article}
%\usepackage{a4wide}\begin{document}
\begin{flushleft}
{\Large The ATLAS Collaboration}

\bigskip

G.~Aad$^{\rm 48}$,
B.~Abbott$^{\rm 110}$,
J.~Abdallah$^{\rm 11}$,
A.A.~Abdelalim$^{\rm 49}$,
A.~Abdesselam$^{\rm 117}$,
O.~Abdinov$^{\rm 10}$,
B.~Abi$^{\rm 111}$,
M.~Abolins$^{\rm 87}$,
H.~Abramowicz$^{\rm 152}$,
H.~Abreu$^{\rm 114}$
% ...
\bigskip

% ...
$^{10}$ Institute of Physics, Azerbaijan Academy of Sciences, Baku, Azerbaijan\\
$^{11}$ Institut de F\'isica d'Altes Energies and Departament de F\'isica de la Universitat Aut\`onoma  de Barcelona and ICREA, Barcelona, Spain\\
% ...
$^{48}$ Fakult\"{a}t f\"{u}r Mathematik und Physik, Albert-Ludwigs-Universit\"{a}t, Freiburg i.Br., Germany\\
$^{49}$ Section de Physique, Universit\'e de Gen\`eve, Geneva, Switzerland\\
% ...
$^{87}$ Department of Physics and Astronomy, Michigan State University, East Lansing MI, United States of America\\
% ...
$^{110}$ Homer L. Dodge Department of Physics and Astronomy, University of Oklahoma, Norman OK, United States of America\\
$^{111}$ Department of Physics, Oklahoma State University, Stillwater OK, United States of America\\
% ...
$^{114}$ LAL, Univ. Paris-Sud and CNRS/IN2P3, Orsay, France\\
% ...
$^{117}$ Department of Physics, Oxford University, Oxford, United Kingdom\\
% ...
$^{152}$ Raymond and Beverly Sackler School of Physics and Astronomy, Tel Aviv University, Tel Aviv, Israel\\
% ...
\end{flushleft}

%\end{document}

\end{document}